\documentclass[twocolumn,preprintnumbers,amsmath,amssymb,showpacs]{revtex4}

\usepackage{graphicx}
\usepackage{dcolumn}
\usepackage{bm}

\begin{document}

\title{Measurements of the Yield Stress in Frictionless Granular Systems}

\author{Ning Xu$^1$} \author{Corey S. O'Hern$^{1,2}$}
\address{$^1$~Department of Mechanical Engineering, Yale University,
New Haven, CT 06520-8284.\\ $^2$~Department of Physics, Yale
University, New Haven, CT 06520-8120.\\ }

\date{\today}

\begin{abstract}

We perform extensive molecular dynamics simulations of 2D frictionless
granular materials to determine whether these systems can be
characterized by a single static yield shear stress.  We consider
boundary-driven planar shear at constant volume and either constant
shear force or constant shear velocity.  Under steady flow conditions,
these two ensembles give similar results for the average shear stress
versus shear velocity.  However, near jamming it is possible that the
shear stress required to initiate shear flow can differ substantially from
the shear stress required to maintain flow.  We perform several
measurements of the shear stress near the initiation and cessation of
flow.  At fixed shear velocity, we measure the average shear stress
$\Sigma_{yv}$ in the limit of zero shear velocity.  At fixed shear
force, we measure the minimum shear stress $\Sigma_{yf}$ required to
maintain steady flow at long times.  We find that in finite-size
systems $\Sigma_{yf} > \Sigma_{yv}$, which implies that there is a
jump discontinuity in the shear velocity from zero to a finite value
when these systems begin flowing at constant shear force.  However, our
simulations show that the difference $\Sigma_{yf} - \Sigma_{yv}$, and
thus the discontinuity in the shear velocity, tend to zero in the
infinite system size limit.  Thus, our results indicate that in the
large system limit, frictionless granular systems are characterized by
a single static yield shear stress.  We also monitor the short-time
response of these systems to applied shear and show that the packing
fraction of the system and shape of the velocity profile can strongly
influence whether or not the shear stress at short times overshoots
the long-time average value.

\end{abstract}

 \maketitle

\section{Introduction}
\label{intro}

The static yield shear stress, or similarly the static shear modulus,
is an important material property that distinguishes solids from
liquids \cite{jamming}.  Solids possess a nonzero static yield shear
stress, while it vanishes for liquids.  Solids are able to resist
applied shear stresses below the yield shear stress, but plastic flow
occurs when shear stresses larger than the yield shear stress are
applied.  In contrast, liquids flow when any finite shear stress is
applied.

Disordered materials such as molecular and colloidal glasses, static
granular materials, and concentrated emulsions also possess a nonzero
yield shear stress. However, it is difficult to determine precisely
the yield shear stress in these amorphous systems since they often
display nonlinear and spatially nonuniform response, for example creep
flow, intermittent dynamics, and shear localization, when shear stress is
applied.  The value of the yield stress in these amorphous systems can
also depend on how it is measured.  For example, the yield shear
stress required to generate steady flow in an originally unsheared
system may differ significantly from a measure of the yield stress
obtained by approaching the static state by slowly decreasing the
shearing velocity.  The yield shear stress may also depend strongly on
how the system was prepared.  For example, it has been shown that the
yield shear stress is sensitive to the age and strain history in
glassy systems \cite{utz} and the construction history
\cite{nowak,vanel} and micro-structural details \cite{losert} in
granular materials.

There have been a number of recent computational investigations of the
transition from static to flowing states in granular and glassy
systems.  For example, measurements of the yield shear stress or
static shear modulus have been conducted as a function of packing
fraction in model foams \cite{durian}, emulsions \cite{mason}, and
frictionless granular materials \cite{ohern} and as a function of
temperature and strain rate in dense Lennard-Jones glasses
\cite{varnik_yield,rottler,rottler_prl}, metallic glasses \cite{fu},
and polymer glasses \cite{he,rottler2}.  However, an important
question that has not been adequately addressed by these previous
studies is whether or not there is a unique measure of the static
yield shear stress in amorphous granular and glassy systems.  Several
studies have pointed out that the shear stress required to initiate
flow can be larger than the shear stress required to prevent slow
shear flows from stopping \cite{varnik_yield,berthier}, but, does this
difference in shear stress persist in the infinite system size limit?
If so, what physical mechanism (for example, force chains in granular
materials \cite{majmudar}) is responsible for the difference?  If not,
how significant are the finite-size effects?

We perform molecular dynamics simulations of frictionless granular
materials subjected to boundary-driven shear at fixed volume to
determine whether or not these simple systems can be characterized by
a single static yield stress in the large system limit.  At constant
shearing velocity, we measure the long-time average shear stress
$\Sigma_{yv}$ in the limit of zero shearing velocity.  We also perform
simulations at fixed shear force and identify the minimum shear stress
$\Sigma_{yf}$ required to maintain steady shear flow at long times.  We
indeed find that $\Sigma_{yf} > \Sigma_{yv}$ at finite system size.
However, the difference tends to zero in the infinite system size
limit.  Thus, we argue that large frictionless granular systems
possess a single static yield shear stress.  In future studies, we
will include static friction to determine whether the gap $\Sigma_{yf}
- \Sigma_{yv}$ remains finite in large frictional granular systems.

We also investigate the short-time response of frictionless
granular systems to applied shear.  Previous studies of sheared glassy
systems \cite{rottler,rottler2,falk} have found that the shear stress in
response to applied shear strains overshoots the long-time average
value at short times.  The shear stress overshoot is often employed as
a dynamic measure of the yield shear stress. In addition, these
studies have found that the size of the overshoot increases with
increasing shear rate and decreasing temperature.  Does the shear
stress overshoot at short times also occur in model granular systems?
Is the shear stress overshoot related to the difference in the
measured values of the yield shear stress $\Sigma_{yf} - \Sigma_{yv}$?
To address these questions, we monitor the short-time response of the
shear stress to applied shear strain over a range of packing fractions
and shear velocities and in systems where we constrain the velocity
profiles to be linear and in systems without such a constraint.  We
find that the packing fraction and shape of the velocity profile
strongly influence the short time response.  In fact, systems near
random close packing with no constraints on the velocity profile do
not possess a shear stress overshoot in the range of shear rate
considered, while systems that are constrained to have linear velocity
profiles do possess an overshoot.
 
\section{Methods}
\label{methods}

In this section, we provide important details of the simulation
methods.  We performed molecular dynamics simulations of frictionless
granular systems in 2D at fixed volume in the presence of an applied
shear stress. The shear stress was applied by moving a top boundary
layer of particles horizontally as a rigid body at either fixed
shearing velocity $u$ or fixed lateral force $F_0$, while the bottom
boundary remained stationary.  We studied systems composed of
$50$-$50$ mixtures of large and small particles with equal mass $m$
and diameter ratio $1.4$.  These bidisperse systems do not crystallize
or segregate under shear \cite{xu_velocity1,xu_velocity2}.

The position ${\vec r}_i$ of each particle $i$ in the bulk was obtained as a
function of time $t$ by solving Newton's equations of motion 
\begin{equation}
\label{eq1}
m \frac{d^2 {\vec r}_i}{dt^2} = {\vec F}_i = \sum_j \left[ F^r_{ij}(r_{ij}) - b
\left( {\vec v}_i - {\vec v_j}\right)\cdot {\hat r}_{ij} \right] {\hat r}_{ij},
\end{equation}
where the sum over $j$ is a sum over the nearest neighbors of particle $i$.
The simple frictionless granular systems considered here interact via
two pairwise forces that act only along the line connecting particle
centers ${\hat r}_{ij}$ and are nonzero only when particles $i$ and
$j$ overlap \cite{luding}.  The first pairwise interaction is the
purely repulsive linear spring force
\begin{equation}
F^r_{ij}(r_{ij}) = \frac{\epsilon}{\sigma_{ij}}
\left(1-\frac{r_{ij}}{\sigma_{ij}} \right),
\label{equ:potential}
\end{equation}
where $\epsilon$ is the characteristic energy scale of the
interaction, $\sigma_{ij}=(\sigma_i + \sigma_j)/2$ is the average
diameter of particles $i$ and $j$, and $r_{ij}$ is their separation.
The second pairwise interaction is dissipative and proportional to
velocity differences along ${\hat r}_{ij}$.  We chose the damping
coefficient $b=0.0375$, which corresponds to a restitution coefficient
$e=0.92$ typical for granular systems.

At constant velocity, the equation of motion for each particle in
the top boundary is trivial, $d^2 x/dt^2 = 0$, subject to $dx/dt =
u$.  At constant shear force $F_0$, each particle in the top boundary
obeys an equation of motion similar to that in Eq.~\ref{eq1}:
\begin{equation}
\label{top}
M \frac{d^2 x}{dt^2} = F_0 + \sum_{i,j} \left( {\vec F}^r(r_{ij}) -
b \left( u - {\vec v}_j \right) \cdot {\hat r}_{ij} \right) {\hat
r}_{ij} \cdot {\hat x},
\end{equation}
where $M$ is the mass of the top boundary.  The second term in
Eq.~\ref{top} is the total horizontal force on particles $i$ in the
top boundary arising from interactions with particles $j$ in the bulk.

The starting configurations were prepared by choosing a packing
fraction $\phi = 0.85$ near random close packing for this system
\cite{dos} and random initial particle positions.  The system was then
allowed to relax at fixed volume to the nearest local energy minimum
using the conjugate gradient method \cite{numrec}.  During the quench,
periodic boundary conditions were implemented in both the $x$- and
$y$-directions.  Following the quench, particles with $y$-coordinates
$y>L_y$ ($y<0$) were chosen to comprise the top (bottom) boundary.
This preparation algorithm created rough and amorphous top and bottom
boundaries, which prevents slip between the bulk and boundary
particles during shear.  After the boundaries are constructed, the
simulation cell was nearly square and contained $N_0$ bulk particles and
$N_b$ particles in the top and bottom boundaries.  Periodic boundary 
conditions in the $x$ direction were employed during shear.

\begin{figure}
\scalebox{0.5}{\includegraphics{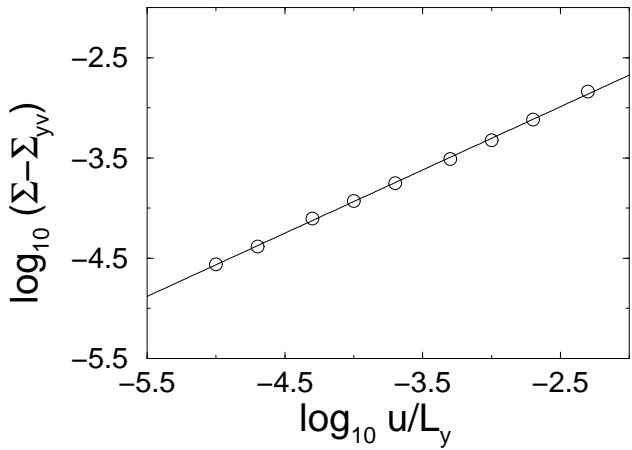}}%
\caption{\label{fig:fig1} Deviation in the shear stress $\Sigma$ from
the shear stress $\Sigma_{yv}$ in the $u \rightarrow 0$ limit versus shear
rate $u/L_y$ for a system with $N_0 \approx 1024$ bulk particles sheared at
constant $u$.  The solid line has slope $0.63$.}
\end{figure}

During the simulations, we calculated the shear stress on the top and
bottom boundaries and in the bulk.  Each of these measurements gave
similar values for the average shear stress, however, the shear stress
fluctuations were much larger on the boundaries as expected.
Therefore, below we focus on measurements of the bulk shear stress
calculated using the virial expression \cite{allen}:
\begin{equation}
\Sigma = - \frac{1}{L_x L_y}\left( \sum_{i=1}^{N_0} \delta v_{xi}\delta
v_{yi} + \frac{1}{2} \sum^{N_0}_{i \ne j} x_{ij}F_{yij}
\right),\label{eq:stress}
\end{equation}
where $\delta {\vec v}_i = {\vec v}_i - \langle {\vec v}_i \rangle$ is
the deviation of the velocity of bulk particle $i$ from the average
velocity $\langle v_i \rangle$ at height $y_i$.

We performed several measurements of the shear stress near the
initiation and cessation of flow.  First, at fixed shearing velocity,
we measured the long-time average shear stress $\Sigma_{yv}$ in the $u
\rightarrow 0$ limit.  Second, at fixed lateral force, we measured the
minimum shear stress $\Sigma = F_0/L_x$ required to maintain steady
shear flow at long times.  For all measurements of the shear stress we
averaged over at least $100$ different initial realizations.  We did
not find large differences in the shear stress response among
different starting configurations.  Also, to assess finite size
effects, we varied the number of particles in the bulk $N_0$ over more
than two orders of magnitude from $N_0 = 32$ to $4096$.  In the
subsequent discussion of results, the small particle diameter
$\sigma$, characteristic energy $\epsilon$, and $\sigma
\sqrt{m/\epsilon}$ were chosen as the units of length, energy, and
time, and all quantities are normalized by these below.

\section{Results}
\label{results}

In this section, we present a number of novel results from our
simulations of frictionless granular materials subjected to
boundary-driven planar shear.  

\begin{figure}
\scalebox{0.5}{\includegraphics{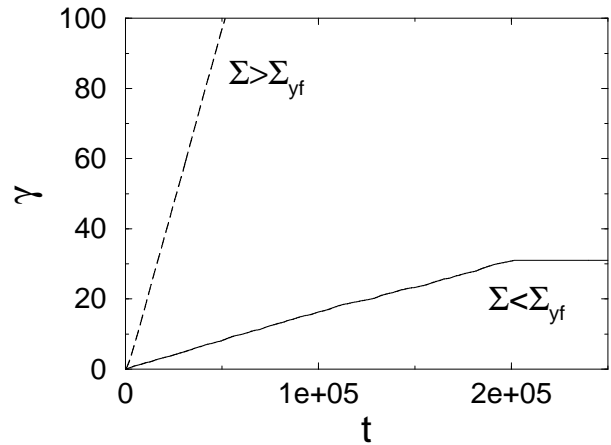}}%
\caption{\label{fig:fig2} Shear strain $\gamma$ versus time $t$ for a
system with $N_0 \approx 1024$ bulk particles sheared at constant
force $F_0 = \Sigma / L_y$.  Two shear stresses $\Sigma = 3.7 \times
10^{-4}$ (solid line) and $\Sigma = 9.7\times 10 ^{-4}$ (dashed line)
are shown.  The smaller value is below and the larger is above the
minimum yield stress $\Sigma_{yf} = 4.4\times 10^{-4}$ required to
maintain steady flow at long times. }
\end{figure}

\subsection{Constant Shearing Velocity}
\label{velocity}

We have measured the average shear stress $\Sigma$ as a function of
shear rate $u/L_y$ in systems sheared at fixed velocity $u$ of the top
boundary.  At each $u$, we began with an unsheared initial
configuration, the system was sheared for a strain of at least $10$,
and then the shear stress was averaged over an additional strain of
$100$.  We have shown in previous studies \cite{xu_velocity2} that at
such large strains these systems are spatially uniform and possess
linear velocity profiles.  We find that the flow curve ($\Sigma$
versus $u/L_y$) for the system obeys the commonly used
phenomenological form \cite{varnik_yield,rottler}
\begin{equation}
\Sigma - \Sigma_{yv} = A_v (u/L_y)^\alpha, \label{flowcurve_fit}
\end{equation}
where $A_v > 0$, $\Sigma_{yv}$ is the shear stress in the $u
\rightarrow 0$ limit, and the power-law exponent $\alpha \approx 0.63$
\cite{remark}. The flow curve for a system with $N_0 \approx 1024$
bulk particles is shown in Fig.~\ref{fig:fig1} and $\Sigma_{yv} =
2.1\times 10^{-4}$ for this system size.  Systems sheared at constant
velocity flow at any nonzero $u$, however, by extrapolating the flow
curve to $u \rightarrow 0$, we can obtain a measure of the yield shear
stress $\Sigma_{yv}$.

\subsection{Constant Shearing Force}
\label{force}

\begin{figure}
\scalebox{0.5}{\includegraphics{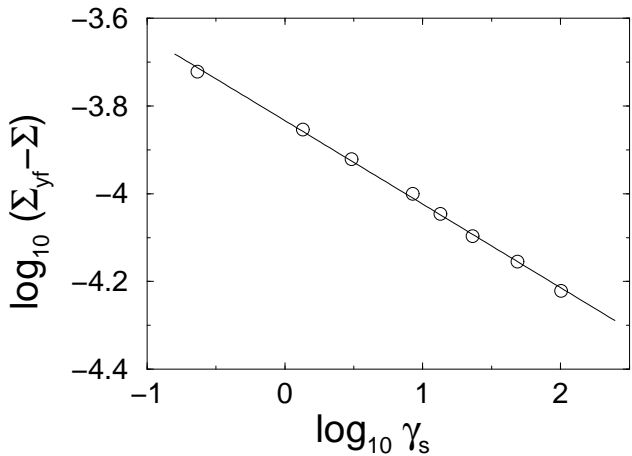}}%
\caption{\label{fig:fig3} The difference between the applied shear
stress $\Sigma$ and the minimum shear stress $\Sigma_{yf}$ required to
maintain shear flow at long times versus the maximum shear strain
$\gamma_s$ obtained.  The solid line has slope $-0.20$.}
\end{figure}

We also studied frictionless granular systems sheared at fixed lateral
force $F_0$.  In this ensemble, granular systems do not flow on long
time scales unless the applied shear stress $\Sigma = F_0/L_x$ exceeds
a shear stress threshold, $\Sigma_{yf}$.  In Fig.~\ref{fig:fig2}, we
show the shear strain $\gamma = x/L_y$, where $x$ is the horizontal
displacement of the top boundary, as a function of time for applied shear
stresses above and below $\Sigma_{yf}$ in a system with $N_0 \approx
1024$ and averaged over $100$ initial realizations.  When $\Sigma >
\Sigma_{yf}$, the shear strain diverges and the system flows at long
times at an average shear rate that is consistent with the flow curve
for the fixed shearing velocity ensemble.  The average shear rate is
given by the slope of strain versus time in Fig.~\ref{fig:fig2}.  When
$\Sigma < \Sigma_{yf}$, the system can flow at short time scales.
However, the system is able to find a configuration that can support
the applied shear stress, and the system stops flowing.  Moreover, the
flow will not resume because dissipation damps the velocity
fluctuations.  As shown in Fig.~\ref{fig:fig3}, the maximum shear
strain $\gamma_s$ that the system attains increases as a power-law
with the applied shear stress $\Sigma$ and diverges as $\Sigma
\rightarrow \Sigma_{yf}$ from below:
\begin{equation}
\Sigma_{yf} - \Sigma = \frac{A_f}{ {\gamma_s}^{\beta} }, \label{sigma_yf_fit}
\end{equation}
where $A_f > 0$, $\Sigma_{yf}$ is the minimum shear stress at which $\gamma_s
\rightarrow \infty$, and the power-law exponent $\beta \approx
0.20$. For $N_0 \approx 1024$ bulk particles, $\Sigma_{yf} = 4.4\times
10^{-4} > \Sigma_{yv}$.

\subsection{System-Size Dependence}
\label{n}

\begin{figure}
\scalebox{0.5}{\includegraphics{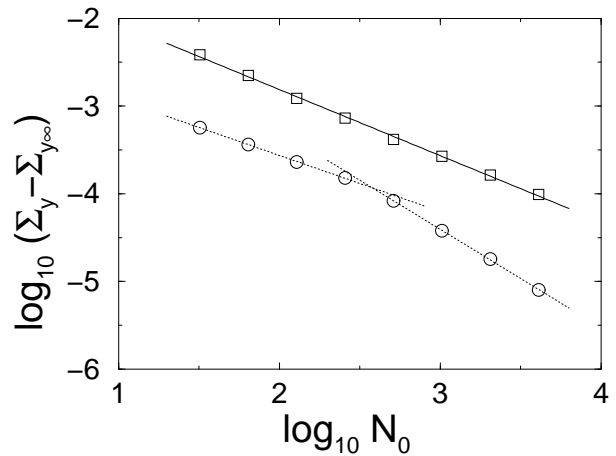}}%
\caption{\label{fig:fig4} The deviation of the yield shear stress
$\Sigma_y$ from its value in the infinite system size limit $\Sigma_{y
\infty}$ as a function of the number of bulk particles $N_0$ in the
constant shear velocity (circles) and force (squares) ensembles.}
\end{figure}

In the previous section, we showed that $\Sigma_{yf} > \Sigma_{yv}$
for systems with $N_0\approx 1024$ bulk particles.  How does the
difference in these two measurements of the yield shear stress depend
on system size?  Does the difference tend to zero for frictionless
granular systems? To answer these questions, we performed measurements
of $\Sigma_{yv}$ and $\Sigma_{yf}$ for systems with $N_0$ in the range
$32$ to $4096$.  For all system sizes, the shear stress obeyed
Eq.~\ref{flowcurve_fit} in the constant shear velocity ensemble and
Eq.~\ref{sigma_yf_fit} in the constant shear force ensemble.  We found
that $\Sigma_{yf} > \Sigma_{yv}$ for all system sizes studied,
however, the difference between these two measures of the yield shear
stress decreased as $N_0 \rightarrow \infty$.  Both measures decreased
with increasing system size and converged to the same value in the
infinite system size limit, $\Sigma_{y \infty} \approx 1.7 \times
10^{-4}$.  In Fig.~\ref{fig:fig4} we show the system-size dependence
of $\Sigma_{yf}$ and $\Sigma_{yv}$.  For example, in the constant 
force ensemble, $\Sigma_{yf} - \Sigma_{y \infty}$
scales as a power law with $N_0$
\begin{equation}
\label{yield_fit}
\Sigma_{y f} - \Sigma_{y \infty} = \frac{B_f}{N_0^{\eta_{f}}}
\end{equation}
over the entire range of system sizes with $B_f > 0$ and $\eta_f
\approx 0.75$.  $\Sigma_{yv}$ has a similar power-law dependence for
small systems, but a more rapid power-law decay with $\eta_v \approx
1.1$ occurs for $N_0 > 300$.  A simple interpretation of $\Sigma_{yf}
\rightarrow \Sigma_{yv}$ is that $\Sigma_{yv}$ measures the average
shear stress, while $\Sigma_{yf}$ is related to the maximum shear
stress in the $u \rightarrow 0$ limit. Since the distribution of shear
stresses becomes a $\delta$-function and the shear fluctuations
vanish, the difference between $\Sigma_{yv}$ and $\Sigma_{yf}$
vanishes in the large system limit for frictionless granular systems.

\subsection{Discontinuity in the Shear Rate}
\label{discontinuity}

\begin{figure}
\scalebox{0.5}{\includegraphics{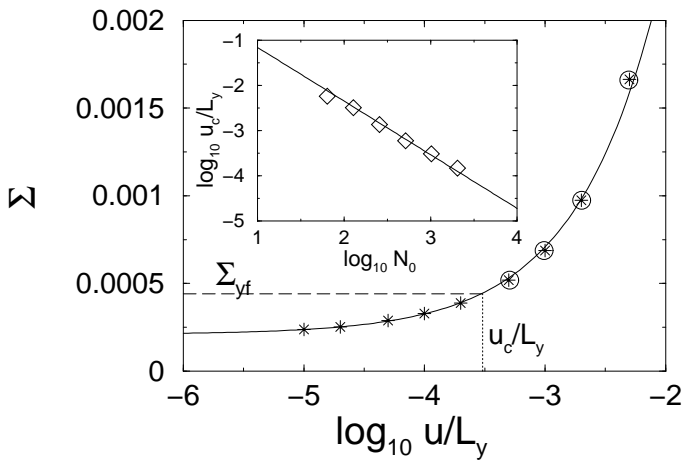}}%
\caption{\label{fig:fig5} Comparison of the flow curves at constant
shear velocity (asterisks) and constant shear force (circles) for
$N_0\approx 1024$.  The solid line is a fit to
Eq.~\ref{flowcurve_fit}.  $u_c/L_y$ is the jump discontinuity in the
boundary shear rate that occurs when the system begins flowing at
constant shear force.  The inset shows the jump discontinuity
$u_c/L_y$ as a function of $N_0$.  The solid line has slope
$\approx -1.2$.}
\end{figure}

In Fig.~\ref{fig:fig5}, we show a comparison of the flow curves,
i.e. shear stress $\Sigma$ versus shear rate $u/L_y$, for systems in
the constant shearing velocity and force ensembles.  For large shear
stresses $\Sigma > \Sigma_{yf}$, there is a correspondence between
shear rate and shear stress in the two ensembles.  However, since
$\Sigma_{yf}$ is larger than the average shear stress $\Sigma_{yv}$ in
the $u \rightarrow 0$ limit at finite system size, there is a jump
discontinuity in the shear rate when the applied shear stress is
increased above $\Sigma_{yf}$.  In fact, several recent experimental
studies have found that foams, emulsions, and granular materials also
display a rate of strain discontinuity when they begin flowing in
response to applied shear stress \cite{dennin1,dennin2,bonn1}.
However, we find that in frictionless granular systems, the jump
discontinuity in shear rate upon initiation of shear flow is a finite
size effect---$u_c$ tends to zero in the infinite system size limit.
The jump discontinuity $u_c(N_0)$ is obtained by solving
$\Sigma_{yv}(u_c,N_0) = \Sigma_{yf}(N_0)$.  Using the scaling
relations in Eqs.~\ref{flowcurve_fit} and~\ref{yield_fit}, we find
that $u_c \sim N_0^{-\eta_f/\alpha} \sim N_0^{-1.2}$, which is
confirmed in the inset to Fig.~\ref{fig:fig5}.  In light of these
results, we advocate further experimental studies of the rate of strain
discontinuities in soft glassy materials, especially in planar shear
cells, to determine whether there are strong finite size effects.

\subsection{Shear Stress Overshoot}
\label{overshoot}

A frequently used measure of the {\it dynamic} yield shear stress is
the shear stress overshoot above the long-time average value
in systems sheared at finite shear rate.  In fact, several recent
computational studies of dense Lennard-Jones glasses have measured the
dependence of the shear stress overshoot on the bath temperature and
imposed shear rate \cite{rottler,rottler2,falk}. In this final
section, we present results from simulations of frictionless granular
systems undergoing planar shear to determine whether a significant
shear stress overshoot occurs on short time scales in the slowly
sheared regime in these systems.

There are several key differences between our current study and
previous investigations of the shear stress overshoot in sheared
glassy systems: 1) we study systems with no constraints on the
velocity profile as well as systems with constraints that enforce a
linear velocity profile $\langle v_{x} \rangle = u y/L_y$ as used
in simulations of glasses \cite{evans}, 2) we study a wide range of
packing fractions from near random close packing at $\phi = 0.85$ to
overcompressed systems at $\phi = 1.1$, and 3) we focus on
dissipative, granular systems, not conservative, thermal systems.  In
the results below, we show that the packing fraction and shape of the
velocity profile strongly influence the short-time response of sheared
granular systems.  We fixed the dissipation in this study, however, the
influence of dissipation will be investigated in a future study
\cite{xu_future}.

\begin{figure}
\scalebox{0.5}{\includegraphics{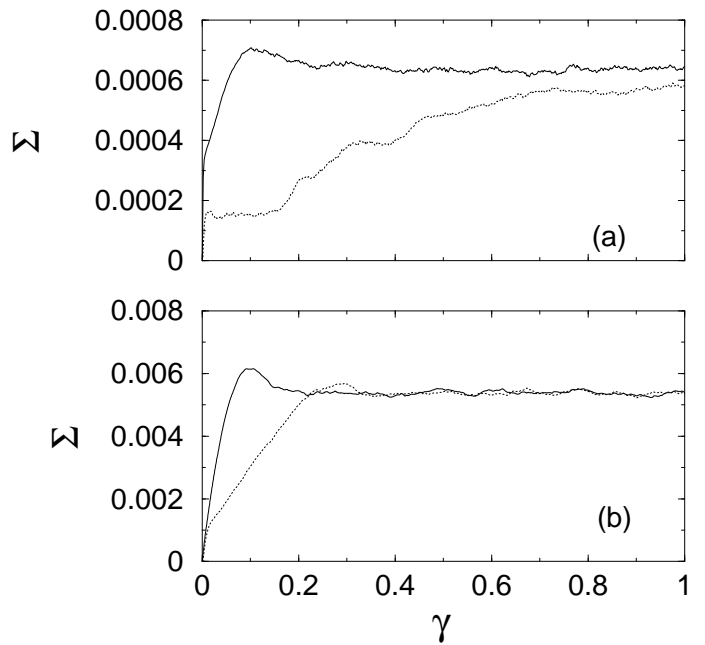}}%
\caption{\label{fig:fig6} Shear stress $\Sigma$ versus shear strain
$\gamma$ for systems with $N_0 \approx 4096$ at (a) $\phi=0.85$ and
(b) $\phi=1.10$ and $u/L_y=0.001$. The solid and dotted lines show
results for systems with and without a constraint that enforces linear
velocity profiles.}
\end{figure}

To study the shear stress overshoot, we measured the shear stress
response of the system at small strains to a slow
applied shear rate $u/L_y = 10^{-4}$.  The shear stress as a function
of shear strain averaged over $100$ independent realizations is shown
in Fig.~\ref{fig:fig6} for a system with $N_0 \approx 4096$.  Note
that in Fig.~\ref{fig:fig6} the shear stress is measured only over
small shear strains up to $\gamma = 1$.  In contrast, the shear stress
was averaged over large shear strains up to $\gamma = 100$ in previous
figures.  

Two striking results are presented in Fig.~\ref{fig:fig6}.  First,
when the velocity profile is unconstrained, the shear stress does {\it
not} overshoot the long-time average shear stress at this shear rate.
The shear stress increases monotonically to the long-time average
value.  The shear strain required to reach the long-time average shear
stress decreases with increasing $\phi$, but is much less than the
shear strain required for the system to possess a linear velocity
profile. See Ref.~\cite{xu_velocity2} for an extensive discussion of
the evolution of the velocity profiles in sheared granular systems.
In contrast, when we constrain the system to possess a linear velocity
profile, a large shear stress overshoot develops.  In addition, the
shear strain required to reach the long-time average shear stress is
$< 0.2$ and roughly independent of packing fraction.  These results
suggest that the shear stress overshoot at short times in glassy and 
granular systems is an artifact of the fact that the velocity field is
constrained to be linear.  At the very least, the constraint
significantly amplifies and speeds up the response of the system to
applied shear.  Second, the packing fraction strongly influences the
height of the shear stress overshoot.  The overshoot for $\phi=0.85$
in Fig.~\ref{fig:fig6} (a) (with the constraint) is nearly a factor of
$10$ smaller than that for $\phi=1.1$ in Fig.~\ref{fig:fig6} (b).
Moreover, if we define $\Delta = (\Sigma_m - \Sigma_p) / \Sigma_p$ as
the relative height of the shear stress overshoot with maximum shear
stress $\Sigma_m$ and shear stress plateau at long times $\Sigma_p$,
$\Delta\approx 0.09$ at $\phi=0.85$ compared to $\Delta \approx 0.15$
at $\phi=1.10$. Both of these findings demonstrate that the shear
stress overshoot is less pronounced in frictionless granular
systems near random close packing with the local dissipation model in
Eq.~\ref{eq1} than in dense Lennard-Jones glasses.

\section{Conclusion}

In this article, we studied model frictionless granular systems near
the initiation and cessation of shear flow using molecular dynamics
simulations of boundary-driven shear flow at constant volume in 2D.
These simulations were performed to address several open questions
concerning the jamming (or unjamming) transition in frictionless
granular systems.  First, we wanted to determine whether these model
systems can be characterized by a single yield shear stress in the
large system limit.  We compared two measures of yield shear stress:
1) the average shear stress $\Sigma_{yv}$ in the limit where the
velocity of the shearing boundary tends to zero and 2) the minimum
shear stress $\Sigma_{yf}$ required for steady shear flow at long
times when a constant force is applied to the shearing boundary.  As
found in previous studies of glassy and granular systems, $\Sigma_{yf}
> \Sigma_{yv}$ in finite-sized systems.  However, these two measures
become identical $\Sigma_{yf} = \Sigma_{yv}$ in the infinite system
size limit in frictionless granular systems.  An important direction
for future research is to determine how the inclusion of static
frictional forces affects these results.  Does the difference
$\Sigma_{yf} - \Sigma_{yv}$ remain finite in the large system limit in
{\it frictional} granular systems?  Recent studies have argued that a
large yield stress difference $\Sigma_{yf} - \Sigma_{yv}$ is
responsible for shear banding---spatially localized velocity
profiles---in dense Lennard-Jones glasses \cite{berthier}.  However,
our results suggest that another mechanism is responsible for shear
banding in frictionless granular systems
\cite{xu_velocity1,xu_velocity2}.  Further studies are required to
determine whether a possible gap $\Sigma_{yf} - \Sigma_{yv} > 0$
contributes to shear banding in large frictional systems.

Another question addressed in this article is whether frictionless granular
systems possess a strain rate discontinuity when they begin flowing at
constant force as has been found in several recent experimental
studies on similar systems \cite{bonn1,dennin1,dennin2}.  We indeed
found a discontinuity in the shear rate upon the initiation of flow,
but the discontinuity is proportional to $\Sigma_{yf} - \Sigma_{yv}$
and therefore tends to zero in the infinite system size limit.  We
recommend further experimental studies in planar shear cells to assess
the finite size effects on the strain rate discontinuity.

We have also investigated the short-time response of the shear stress
of the system when the shearing boundary is driven at fixed velocity.
It is well-known for glassy systems that the shear stress at short
times can overshoot the long-time average value when these systems are
sheared at finite shear rate, and the height of the overshoot is often
used as a measure of the dynamic yield shear stress.  We found several
novel results for the short-time response.  First, the shape of the
velocity profile strongly influences the shear stress at short times.
When our systems did not have a constraint imposed on the velocity
profile, we did not observe an overshoot in the shear stress.  In
contrast, when a linear velocity profile was enforced, a strong shear
stress overshoot occurred.  Second, the height
of the overshoot (at least in systems with linear velocity profiles)
decreases with decreasing packing fraction.  For example, the height
of the peak in shear stress is a factor of $10$ smaller near random
close than in an overcompressed system at $\phi=1.1$.  These results
point out that there is not a significant shear stress overshoot in
frictionless granular systems---in contrast to dense Lennard-Jones
glasses.

\subparagraph{Acknowledgments} 

Financial support from NSF grant numbers CTS-0456703 (CSO) and
DMR-0448838 (NX,CSO) is gratefully acknowledged.  We also thank Yale's
High Performance Computing Center for generous amounts of computer
time.

\end{document}